\documentclass[10pt]{iopart}
\usepackage[T1]{fontenc}
\usepackage[british]{babel}
\usepackage[utf8]{inputenc}
\usepackage{graphicx}
\usepackage{iopams}

\begin{document}

\title{Recombination of Open-f-Shell Tungsten Ions}

\author{C Krantz$^1\footnote{Present address: Marburg Ion-Beam Therapy Centre, 35043 Marburg, Germany}$, N R Badnell$^2$, A Müller$^3$, \\S Schippers$^4$, A Wolf$^1$}

\address{$^1$ Max-Planck-Institut für Kernphysik, 69117 Heidelberg, Germany}
\address{$^2$ Department of Physics, University of Strathclyde, Glasgow G4 0NG, United Kingdom}
\address{$^3$ Institut für Atom- und Molekülphysik, Justus-Liebig-Universität Gießen, 35392 Giessen, Germany}
\address{$^4$ I.\ Physikalisches Institut, Justus-Liebig-Universität Gießen, 35392 Giessen, Germany}
\ead{claude.krantz@med.uni-heidelberg.de}

\vspace{10pt}
\begin{indented}
\item[]\today
\end{indented}

\begin{abstract}
  We review experimental and theoretical efforts aimed at a detailed understanding of the recombination of electrons with highly-charged tungsten ions characterised by an open 4$f$ sub-shell. Highly-charged tungsten occurs as a plasma contaminant in ITER-like tokamak experiments, where it acts as an unwanted cooling agent. Modelling of the charge state populations in a plasma requires reliable thermal rate coefficients for charge-changing electron collisions. The electron recombination of medium-charged tungsten species with open 4$f$ sub-shells is especially challenging to compute reliably. Storage-ring experiments have been conducted that yielded recombination rate coefficients at high energy resolution and well-understood systematics. Significant deviations compared to simplified, but prevalent, computational models have been found. A new class of {\it ab-initio} numerical calculations has been developed that provides reliable predictions of the total plasma recombination rate coefficients for these ions. 
\end{abstract}

\submitto{\jpb}
\maketitle
\ioptwocol
\section{Introduction}
Control of contamination by atoms of high nuclear charge ($Z$) is an important aspect in the preparation of technical fusion plasmas. The presence of heavy elements in the otherwise light (low-$Z$) element plasma causes bremsstrahlung as well as X-ray and VUV line emission. Due to the low column density of the plasma, this radiation is effectively not re-absorbed and, hence, drains energy from the plasma. In nuclear fusion reactors, this radiation-related energy loss can ultimately prohibit ignition of the plasma \cite{puetterich2010}.\par
The wall material of the plasma-containing vacuum vessel constitutes a natural source of plasma contamination. Surface bombardment leads to sputtering of atoms off the exposed surfaces. Initially neutral, the atoms are collisionally ionised while migrating to the hotter areas of the plasma. Their final charge states are determined by balance between electron-impact ionisation and electron-ion recombination.\par
To minimise impurity-related radiation losses, most of the plasma-facing first-wall surface of the upcoming ITER facility will consist of beryllium tiles \cite{matthews2007,merola2014}. However, a different armour material is necessary for the so-called divertor sections of the tokamak, which are exposed to the most intense plasma bombardment. Here, in spite of its high nuclear charge, tungsten has eventually been chosen over carbon, as it combines the ability to withstand extreme thermal loads with a low tritium retention factor \cite{merola2014}. The importance assigned to the resulting tungsten contaminations to be expected at ITER is documented by the fact that several precursor experiments (JET, ASDEX-Upgrade, JT60-U) have been equipped with tungsten wall armour in order to study its effects on the tokamak plasma \cite{matthews2007,neu2009,nakano2009}.\par
Reliable interpretation of the experiments and predictions for future set-ups require accurate atomic data on all relevant charge states of tungsten ions. Besides the atomic level structures, rate coefficients for electron-impact ionisation and electron-ion recombination in the temperature range relevant to fusion plasmas are necessary for modelling the equilibrium charge state populations. All-in-all, the amount of data is much beyond what dedicated experimental precursor studies could realistically deliver on their own. The bulk of the data has to be provided by reliable calculations, best backed by selected experimental benchmarks.\par 
The low-to-intermediate charge states, characterised by an open 4$f$ sub-shell, are especially challenging in this respect. The charge state of these ions is high enough to render their production in the laboratory difficult, and the complexity of their electronic configurations drives state-of-the art atomic structure computations to their limits.\par 
In this article, we review a series of experiments and calculations on electron recombination of tungsten ions in the range of charge states from W$^{18+}$ to W$^{21+}$. In all cases, both the experimental and theoretical analysis revealed significant discrepancies to the predicted recombination rate coefficients \cite{puetterich2008,foster2008} held in the ADAS database \cite{adas}, which are widely used in modelling tokamak plasmas. Beyond its direct implications for the fusion plasma application, the work may also be seen as a prototype study for an entire class of electron-ion recombination reactions involving extremely complex atomic shell structures.\par
Although the work we review here is the result of parallel evolution and interplay of theory and experiment, we split this article into an `experimental' and a `theoretical' part, hoping to improve readability. Following an overview of the wider scope of the research in Sect.~\ref{overview}, Sects.\ \ref{expsect} and \ref{expdata} present the experimental method and the resulting experimental data sets. Sections \ref{theorychapter} and \ref{discussion} describe the theoretical developments that took place within the time frame of our studies, and discuss the present state-of-the-art of the computational models as compared to the experimental benchmarks.   
\section{Related work\label{overview}}
Due to their importance to the fusion community, electron collision processes of highly-charged tungsten ions have drawn quite some attention in the past years. Besides {\it in-situ} spectroscopy of tungsten ions in actual tokamak plasmas, theoretical and experimental studies of selected electron collision processes have been conducted. The work we review, focussing specifically on recombination of 4$f$-sub-shell tungsten, is itself part of a broader study targeting also photo-ionisation and electron-impact ionisation of a variety of tungsten charge states \cite{mueller2015}.\par
Spectroscopic studies of plasma pollution by tungsten impurities have been done extensively at the ASDEX-Upgrade experiment \cite{neu2009}. When modelling VUV and X-ray emission spectra of tungsten ions, Pütterich et al.\ \cite{puetterich2008}  found disagreements between their observations and predication obtained from the widely used ADAS atomic data base \cite{adas}. They noted that improved agreement between observation and theory could be obtained by deliberately increasing the recombination rate coefficients for the low-to-medium charge states in the range between W$^{20+}$ and W$^{40+}$ by up to factors of two, thus changing the ionisation balance and the dominant radiating charge-state at a given temperature \cite{puetterich2008,adas}. The discrepancy between prevalent numerical models of charge-state populations and experimental results in the case of tungsten has been discussed on several occasions by the plasma modelling community \cite{fontes2009,ralchenko2009,chung2013}.\par
For non-bare ions, recombination with free electrons happens primarily via resonant-capture into an auto-ionising level, followed by radiative stabilization to a truly bound one -- most prominently via dielectronic recombination (DR) \cite{burgess1964,shore1969}. Reliable predictions of the total recombination rate coefficients must account for all significant auto-ionising levels that are accessible in the resonant-capture step of the reaction. Hence, they must be based on detailed atomic structure calculations. Especially in the low-temperature region, neglecting even part of the resonant reaction channels can lead to significant underestimation of the total calculated recombination yields. This had been found previously, e.g., in the case of astrophysically relevant charge states of iron with open $L$ or $M$ shells \cite{savin1997,savin2003,lestinsky2009,badnell2006b}. Similar deficiencies were to be expected also in the model calculations for tungsten ions -- atomic systems of significantly greater complexity.\par
Prior to the studies described here, theoretical work on tungsten recombination was scarce, and precision experimental data was practically non-existing. In the absence of more reliable data, average atom (ADPAK) or the semi-empirical Burgess formalisms were used \cite{puetterich2008,foster2008,burgess1965} to calculate the recombination rate coefficients. Level-to-level calculations of tungsten recombination, based on atomic structure theory, were available only for very high charge states \cite{behar1999,safranova2011} that are of limited relevance for the fusion application. Nevertheless, even in a system as highly charged as W$^{56+}$, Peleg et al.\ noted that the low-temperature recombination rate coefficient of the ADAS database fell short of their detailed calculation \cite{peleg1998}. Recently, Kwon and Lee published detailed work on W$^{46+}$, W$^{45+}$, and W$^{44+}$ where they noted similar discrepancies \cite{kwon2016,kwon2016b}.\par
The lower charge states of tungsten, characterised by open $d$ and $f$ sub-shell configurations, were long out of reach of detailed calculations due to the immense complexity of their atomic structure.  For these charge states, the ADAS database only features rate coefficients based on said average-atom and Burgess formalisms. Although this approximate treatment of recombination is applicable to high temperature plasmas only, in lack of better options, the tokamak community widely used (and still uses) the ADAS data also for models of the relatively low-temperature fusion plasma edge. Only in 2010, an initial effort to improve the available recombination rate coefficients also for the lower charge states of tungsten was conducted by Ballance et al., when they published a level-to-level calculation for W$^{35+}$, an ion with open $d$ shell \cite{ballance2010}.\par
During the past few years, we have conducted and, in part, published our experimental measurements on the four neighbouring open-$f$-shell charge states W$^{18+}$ to W$^{21+}$ \cite{schippers2011,spruck2014,badnell2016}. Partly as a result of the newly available experimental data, updated theoretical calculations for a variety of tungsten ions in the range of open $d$ and $f$ sub-shells have emerged. Besides our own theoretical work \cite{badnell2012,badnell2016}, that we review in Sect.~\ref{theorychapter}, Dzuba et al.\ independently provided innovative theoretical models that could be benchmarked against our measurements \cite{dzuba2012,dzuba2013}. Li et al.\ and Safranova et al.\ published updated theoretical work on W$^{29+}$, W$^{28+}$, and W$^{27+}$ \cite{li2012,li2016,safranova2011b}.\par 
Recently, a new initiative {\it The Tungsten Project} has begun to report total and partial final-state resolved DR and RR rate coefficients obtained with the \textsc{Autostructure} code \cite{autostructure}, with the ultimate goal of covering the whole isonuclear sequence consistently. Results to date cover W$^{74+}$--W$^{56+}$ \cite{preval2016a}, W$^{55+}$--W$^{38+}$ \cite{preval2016b} and W$^{37+}$--W$^{28+}$ \cite{preval2016c}. 
\section{Experimental method\label{expsect}}
The experimental part of the work we review here consisted of accurate measurements of the recombination rate coefficients of the four tungsten charge states from W$^{18+}$ to W$^{21+}$ using the co-linear merged-beams technique at a heavy-ion storage ring. The method, as laid out in the following, yields datasets that are systematically quite robust and straight forward to interpret, but involves a rather complex experimental set-up that we outline in this section.\par
\subsection{Merged beams}
Generally, in the merged-beams technique, two fast-propagating particle beams are overlapped co-linearly in a common transport line \cite{phaneuf1999}. We refer to the latter as the `target' or `interaction' region, as it is the volume in which interaction among the two partner beams occurs. The merging and de-merging of the two beams is possible if they are characterised by different ion-optical rigidities. In the case of ion-electron interaction, this condition is always fulfilled. In contrast to a collider or crossed-beam geometry, the co-linear merged-beams set-up allows almost arbitrarily low collision energies. In the non-relativistic limit\footnote{For clarity, we restrict the discussion to the classical case here. The equivalent relativistically correct equations for the collision kinematics can be found in specialised publications \cite{kilgus1992,kieslich2004}.}, let $E_\mathrm{e}$ and $E_\mathrm{i}$ be the longitudinal kinetic energies of the electron (index e) and ion (index i) beams in the laboratory frame, and let $m_\mathrm{e}$ and $m_\mathrm{i}$ be the respective particle masses. Let us first assume that the particles to have no velocity components perpendicular to their common beam axis. As $m_\mathrm{e} \ll m_\mathrm{i}$, one sees that the collision energy is approximately given by
\begin{equation}
 \label{Ecoll_nonrel}
 E_\mathrm{coll} = \frac{1}{2} m^* v_\mathrm{coll}^2 \approx \left( \sqrt{E_\mathrm{e}} - \sqrt{\frac{m_\mathrm{e}}{m_\mathrm{i}}E_\mathrm{i}} \right)^2 = E_\mathrm{d} \ .
\end{equation}
Therein, $m^* \approx m_\mathrm{e}$ is the reduced mass of the collision. $v_\mathrm{coll}$ is the true collision velocity between electrons and ions, whereas $E_\mathrm{d}$ is the experimentally defined `detuning' energy between the two merged beams. The longitudinal beam energies $E_\mathrm{e}$ and $E_\mathrm{i}$ can be chosen such that $E_\mathrm{d}$ vanishes, although $E_\mathrm{e}$ and $E_\mathrm{i}$, themselves, can be quite high. If the beams are perfectly parallel and monochromatic, $E_\mathrm{d}$ and $E_\mathrm{coll}$ are practically identical. In presence of non-zero energetic or angular spreads in the two beams, as discussed below, $E_\mathrm{coll}$ will slightly deviate from $E_\mathrm{d}$, but may still be very small for advanced set-ups.\par
In the following we consider reactions of a beam of tungsten ions W$^{q+}$ (in charge state $q$) with a co-linear electron beam. Recombination leads to a changed charge state $q^\prime = q-1$ of the ion. The capture of the electron may proceed directly, by radiative recombination (RR), or resonantly, via intermediate auto-ionising states, as in dielectronic recombination (DR) \cite{wolf2000}. Independent of the details of the recombination mechanism, the net reaction
\begin{equation}
 \mathrm{W}^{q+} + e^- \rightarrow \mathrm{W}^{(q-1)+} + \sum_k\gamma_k\ ,
\end{equation}
yields a number of photons $\gamma_k$ as well as a \emph{daughter beam} of W$^{(q-1)+}$ ions that leave the interaction region. In the laboratory frame, the energy $E_\mathrm{i}$ of the heavy-ion beam is so large compared to the collision energy $E_\mathrm{coll}$ that the momentum transfer related to the recombination process can be neglected. Thus the product W$^{(q-1)+}$ can be assumed to leave the target with the unmodified momentum of its parent W$^{q+}$.\par 
While, also in merged-beams experiments, observation of the emitted photon spectrum $\sum_k\,\gamma_k$ is possible \cite{stoehlker2000}, detection of the massive product W$^{(q-1)+}$ alone is already sufficient to derive the recombination rate coefficient of the collision \cite{wolf2000}. The daughter beam leaving the target can be easily separated from the parent beam using an analysing magnetic field, and recorded using a single-ion detection system. Thanks to the high energy and the forward projection in the laboratory frame, one can reach detection efficiencies near unity, even using relatively small and simple particle detectors \cite{rinn1982,spruck2015}.\par
From the measured product formation rate $R$, the recombination rate coefficient $\alpha$ can be derived. $\alpha$ is the product of the collision velocity $v_\mathrm{coll}$ and the recombination cross-section section $\sigma$, which is itself velocity-dependent:
\begin{equation}
 \label{singlepassrate}
 \alpha = \sigma(v_\mathrm{coll})v_\mathrm{coll} = R\,v_e v_\mathrm{i}/\Omega = R\, v_e v_\mathrm{i} \left/\int_V j_\mathrm{e} j_\mathrm{i} \mathrm{d}V \right. .
\end{equation}
$v_\mathrm{e}$ and $v_\mathrm{i}$ are the velocities of the electron and ion beams in the laboratory frame, $j_\mathrm{e}$ and $j_\mathrm{i}$ are the corresponding particle flux densities, and $\Omega$ is the overlap integral \cite{phaneuf1999}.\par
Generally, computation of $\Omega$ requires knowledge of the spatial distribution of the flux densities $j_\mathrm{e}$ and $j_\mathrm{i}$ in the target volume $V$, which can be difficult to obtain with good precision. Also, in a real-life experiment, the collision energy $E_\mathrm{coll}$ has contributions from the velocity spreads of the two particle beams in addition to the detuning energy $E_\mathrm{d}$. The actual collision velocities $v_\mathrm{coll}$ are, hence, distributed close to the velocity $v_\mathrm{d}=\sqrt{2 E_\mathrm{d}/m_\mathrm{e}}$ that results from the detuning energy $E_\mathrm{d}$ (cf.\ Eq.~\ref{Ecoll_nonrel}). Correspondingly, the measured recombination rate coefficient $\tilde{\alpha}$ is the convolution of the recombination rate coefficient with that collision velocity distribution around $v_\mathrm{d}$: $\tilde{\alpha}{(v_\mathrm{d})} = \left\langle \sigma(v_\mathrm{coll})v_\mathrm{coll} \right\rangle$.\par
\subsection{Recombination in electron cooler storage rings}
Practically all ion storage rings in operation descended from the first strong-focussing proton synchrotrons, developed in the 1950s as main acceleration stages for nuclear and particle physics experiments \cite{courant1952}. On that basis, specialised machines have been conceived, geared towards DC (`coasting') beam storage of medium-energy ($\sim 1$--100~MeV/u) heavy ions in a wide range of charge states \cite{mullerwolf1997}. This became possible as advanced vacuum systems emerged, which reach residual gas densities of the order of $10^{-11}$~mbar in the storage ring beam line. Examples of such machines are the ESR at GSI/FAIR, CRYRING of the Manne-Siegbahn Laboratory in Stockholm (now also at GSI/FAIR), the TARN II storage ring in Tokyo, ASTRID in Aarhus, and the TSR in Heidelberg \cite{franzke1987,abrahamsson1993,tanabe1998,stensgaard1988,habs1991}. Compared to a single-pass ion beam line, a storage ring combined with an in-ring target section comes with several advantages. \par
\begin{figure}[tb]
 \centering
 \includegraphics[width=8.3cm]{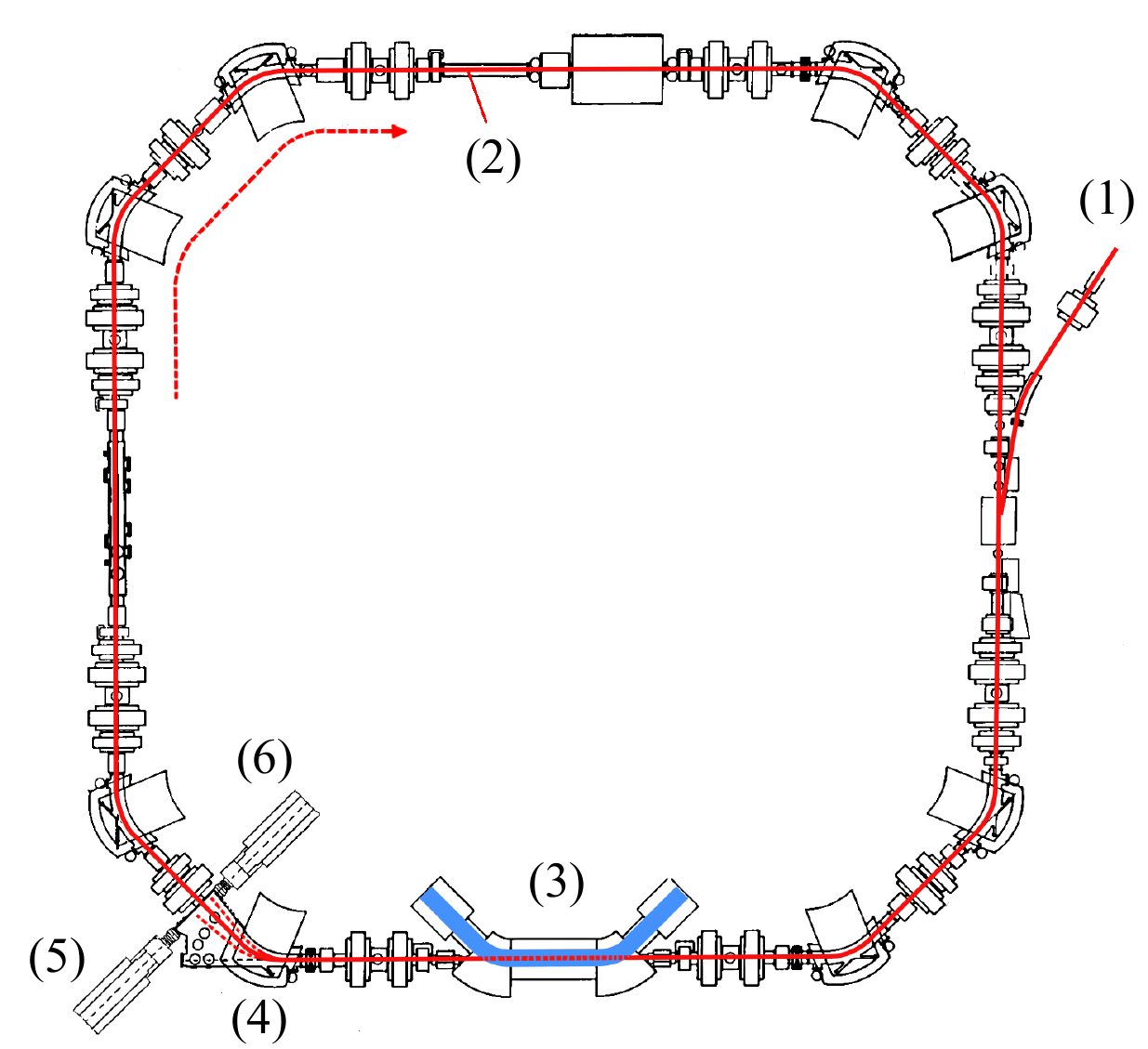}
 \caption{Schematic drawing of the ion storage ring TSR (view from top). Ions are injected into the ring from a tandem accelerator (1, not shown). After multi-turn-injection, a lattice of bending and focussing magnets keeps the ions on a 55.4~m long closed orbit (2) at constant velocity. The arrow indicates the sense of revolution of the ions. The electron cooler (3) overlaps with the ion trajectory, creating an electron-ion interaction section of approximately 1.5~m. The bending dipole magnet (4) following the cooler acts as charge-to-mass analyser and separates daughter particles, created in the interaction region of the cooler, from the stored parent beam. A single-ion detector on the outside of the storage orbit intercepts the path of the recombination products (5), whereas a similar device on the inside of the ring can detect ionisation products (6).\label{tsrfig}}
\end{figure}
First, the storage ring allows the accumulation of ions, e.g.\ by multi-turn injection, or by stacking of several independent injection pulses from the accelerator \cite{grieser2012}. This yields a beam of higher current intensity than would be available from the ion source directly, potentially improving the signal-to-background ratio in the experiments.\par
Second, by storing the ions for many revolutions before starting the experiment, the effective time-of-flight of the ions from the accelerator to the target can be made almost arbitrarily long -- limited only by the mean free path of the ions in the storage ring vacuum. This is especially important if unwanted metastable levels are populated upon production of the ions in the source or accelerator. Often, mere storage of the ions for a few seconds before starting the data acquisition is sufficient to make sure that the reactant beam is characterised by a well-known initial-state population.\par
Last, active beam shaping techniques can be applied to the stored ions prior to undertaking the experiment. Among those are the various methods of beam cooling, of which here only \emph{electron cooling} is of importance.\par
An electron cooler \cite{liesen1987} consists of an intense electron beam overlapping the stored ion beam at matched velocity in a straight section of the storage ring. Naturally, an electron cooler can thus also serve as an in-ring merged-beams electron \emph{target}.\par
Whereas the cold electrons pass by the interaction section only once, the stored ions interact with the electron medium repeatedly at each revolution in the ring. Any difference in velocity between a particular ion and the electron beam results in repeated Coulomb scattering between the ion and the electrons, whose momentum transfers sum-up to an overall friction force, opposing the ion motion relative to the electron medium. To a good approximation, the process can be described analogously to the Bragg stopping of ions in matter \cite{poth1990}. In combination with the dynamics of a coasting beam in the storage ring, this stopping force leads to a shrinking of the beam in five phase space dimensions. In the co-moving frame of the beams, the temperature of the cooler electrons can be of the order of $\sim$10~K for advanced set-ups \cite{danared2000,krantz2009}. Due to competing heating effects, the equilibrium ion-beam temperature is normally somewhat higher.\par
Once the electron cooling process is complete, the cooler beam can be used as a target to probe ion recombination as described above -- either by keeping both beams at tuned velocities ($E_\mathrm{d} = 0$~eV in Eq.~\ref{Ecoll_nonrel}) in the laboratory, or by inducing a certain collision velocity $v_\mathrm{d}$ via a shift of the energy of the cooler electrons away from the cooling condition ($E_\mathrm{d} \ne 0$~eV). The detuning of the energy also results in a renewed friction force experienced by the ions which, eventually, would accelerate the ions towards the velocity of the electron beam. This unwanted effect is referred to as `dragging'. One way of mitigation is fast ($\sim$~ms) switching of the electron energy between the `cooling' and `probing' velocities \cite{kieslich2004}.\par 
The cooling process strongly reduces the width of the ion beam, to values that are typically much smaller than the diameter of the electron beam. This has important and very convenient consequences regarding Eq.~\ref{singlepassrate}: The integration volume can now be chosen very narrow around the ion beam of total electric current $I$. Across this small volume, the flux density $j_\mathrm{e}$ of the much wider electron beam can be considered constant. Hence, the overlap integral from Eq.~\ref{singlepassrate} simplifies to $\Omega \approx j_\mathrm{e}\,L\,I/(q\,e)$, where $L$ is the length of the target. With the ion charge state $q$, $I/(q\,e\,v_\mathrm{i})$ is the linear ion density in the beam, which, using the total number of stored ions $N_\mathrm{i}$ and the circumference of the ring $C$, can be expressed as $N_\mathrm{i}/C$. $j_\mathrm{e}/v_\mathrm{e}$ is equal to the electron number density, which we denote by $n_\mathrm{e}$. Still assuming non-relativistic motion, we thus obtain 
\begin{equation}
  \tilde{\alpha}(v_d) = \langle\sigma(v_\mathrm{coll}) v_\mathrm{coll}\rangle = \frac{R}{n_\mathrm{e}\,N_\mathrm{i}}\frac{C}{L} 
  \label{mbrrc}
\end{equation}
by substituting the above quantities in Eq.~\ref{singlepassrate}.\par
Eq.~\ref{mbrrc} gives the recombination rate coefficient from a set of known or directly measurable quantities: $n_\mathrm{e}$ can be derived from the electron cooler beam current, profile, and energy, $N_i$ is calculated, e.g., from the ion current in the storage ring, $R$ is the directly measured W$^{(q-1)+}$ product rate, and $C/L$ is known from the geometry of the set-up \cite{kilgus1992}.\par
In order to derive the recombination cross section $\sigma$ from the measured recombination rate coefficient $\tilde{\alpha}$, the distribution of $v_\mathrm{coll}$ around $v_d$ needs to be known. The most common variant of electron coolers use a magnetic guiding field for merging and de-merging the electron beam with the ion trajectory. The guiding magnetic field is weak compared to the magnetic rigidity of the ion beam, but strong enough to force the electrons onto helical trajectories that closely follow the magnetic field lines. The longitudinal and transverse degrees of freedom are therefore efficiently decoupled in the motion of the electrons, and -- in the centre of mass frame of the electron-ion collisions -- the electron beam can be characterised by two different temperatures, $T_{\vert\vert}$ and $T_{\perp}$ \cite{danared1997}. As, after electron cooling, the temperatures of both beams are similar, the velocity spread of the much heavier ions is significantly smaller than that of the electrons. The distribution of $v_\mathrm{coll}$ is therefore often identified with the velocity distribution of the electrons.\par 
\section{Recombination of tungsten ions at the TSR\label{expdata}}
In the reviewed work, electron recombination of four neighbouring charge states in the range of open-$f$-shell tungsten has been measured using the storage-ring electron target method. So far, we have published detailed discussions of the experiments and data analysis for W$^{20+}$ (Schippers et al.\ \cite{schippers2011}, Badnell et al.\ \cite{badnell2012}), W$^{18+}$ (Spruck et al.\ \cite{spruck2014}), and W$^{19+}$ (Badnell et al.\ \cite{badnell2016}). For W$^{21+}$, an initial analysis of the experimental data has been provided by Spruck \cite{spruckphd}, while a dedicated article on that charge state is expected in the near future.\par 
All measurements reviewed here have been done at the now decommissioned heavy-ion accelerator facility of the Max Planck Institute for Nuclear Physics. The latter consisted of a set of three injection accelerators (a single-stage Van-de-Graaff accelerator, a larger Tandem-Van-de-Graaff, and an RFQ structure), followed by a radio-frequency post-accelerator and the heavy-ion storage ring TSR (cf.\ Fig.~\ref{tsrfig}) \cite{repnow2002}. As one of the last major measurement campaigns, the experiments on tungsten ions -- difficult in terms of beam manipulation -- once more demonstrated the versatility of that proven set-up.
\subsection{Data taking}
The four tungsten ions in the range of charge states from W$^{18+}$ to W$^{21+}$ were separately produced and injected into the electron-cooler storage ring TSR. As injector, the 12-MV Tandem-Van-de-Graaff accelerator was used. Stripping of a precursor beam of WC$^-$ produced the desired ion charge states at final kinetic energies between 166~MeV (0.90~MeV/u, W$^{18+}$) and 208~MeV (1.13~MeV/u, W$^{21+}$).\par
The experiments were challenging compared to similar measurements on lighter ions. The production efficiency of the desired charge states in the tandem accelerator was low. In spite of the capability of the storage ring to accumulate the ions by multi-turn injection, the stored ion current was very weak by TSR standards -- estimated to have been of the order of $\sim$ 1~nA directly after injection \cite{schippers2011,spruck2014}. Hence, during the experiments the stored beam intensity was well below the sensitivity limit of the DC transformer normally used for measuring the ion current.\par
Usually, the number of parent ions $N_\mathrm{i}$ -- required to calculate the absolute recombination rate coefficient from the measured product yield $R$ (c.f.\ Eq.~\ref{mbrrc}) -- is derived from that direct current measurement. As this was not possible with acceptable precision, a different approach was chosen, as first described by Pedersen et al.\ in the scope of recombination experiments on molecular ions, where very weak stored beams are common \cite{pedersen2005}. The method relies on measurement of a proxy signal of the stored-ion current, calibrated against a single absolute measurement of the rate coefficient by controlled variation of the target density.\par 
In the case of tungsten, collisional ionisation was observed in addition to the recombination signal, using a second product particle detector (position 6 in Fig.~\ref{tsrfig}). As the vacuum pressure in the target section was practically constant over the relevant time scales, and as collisional ionisation in the electron target does not occur in the studied energy range, the ionisation signal was a sensitive proxy for the stored ion number. It was calibrated against an independent direct measurement of the recombination rate coefficient at matched electron and ion velocities ($E_\mathrm{d}=0$~eV).\par 
The calibration measurements made use of the fact that the storage lifetime of the ions in the TSR -- normally in the order of tens of seconds -- was reduced drastically (to values around 1~s) as soon as the cooler electron beam was switched on \cite{schippers2011,spruck2014}. At a known electron density $n_\mathrm{e}$, this allowed us to derive $\tilde{\alpha}(0~\mathrm{eV})$ directly from the respective measured beam lifetimes $\tau_\mathrm{off}$ and $\tau_\mathrm{on}$, via the equation \cite{novotny2012}
\begin{equation}
  \label{alphanorm}
  \tilde{\alpha}(0~\mathrm{eV}) = \frac{ \tau^{-1}_\mathrm{on} - \tau^{-1}_\mathrm{off} }{n_\mathrm{e}\,L/C } \ .
\end{equation}\par
The strong dependence of the storage lifetime on the electron density was an early indication of an unusually high recombination rate coefficient of the tungsten ions at low collision energy. It also required a trade-off between beam cooling and storage efficiency. In most of the experiments, the fast switching between the `probing' and `cooling' energy of the electron beam, as described above, was not done. Instead, only a short electron cooling of 1--2~s was performed during and after ion injection. Then a range of collision velocities was probed in a single, fast scan. While this method allowed us to reach acceptable counting statistics within the available beamtime, it is also believed to have resulted in imperfect ion-beam cooling \cite{spruck2014}.\par
Recombination was measured at collision energies ranging from 0~eV to several hundred eV. For practical reasons, the full energy range of interest could not be scanned during a single cycle of the storage ring. Instead, shorter, mutually overlapping energy intervals were measured and subsequently combined into a common dataset for each charge state. As the absolute calibration of the recombination rate coefficients described above was done at zero collision energy, and as the counting statistics is highest in the low-energy region, the total experimental uncertainty rises with collision energy. At the lowest energies, the total relative uncertainty of the recombination rate coefficient is derived to be  below 10\,\%, while at the upper ends of the energy scales, relative errors of the order of 100\,\% have to be expected \cite{spruck2014,badnell2016}.\par
\subsection{Experimental findings}
\begin{figure}[tb]
 \centering
 \includegraphics[width=8.3cm]{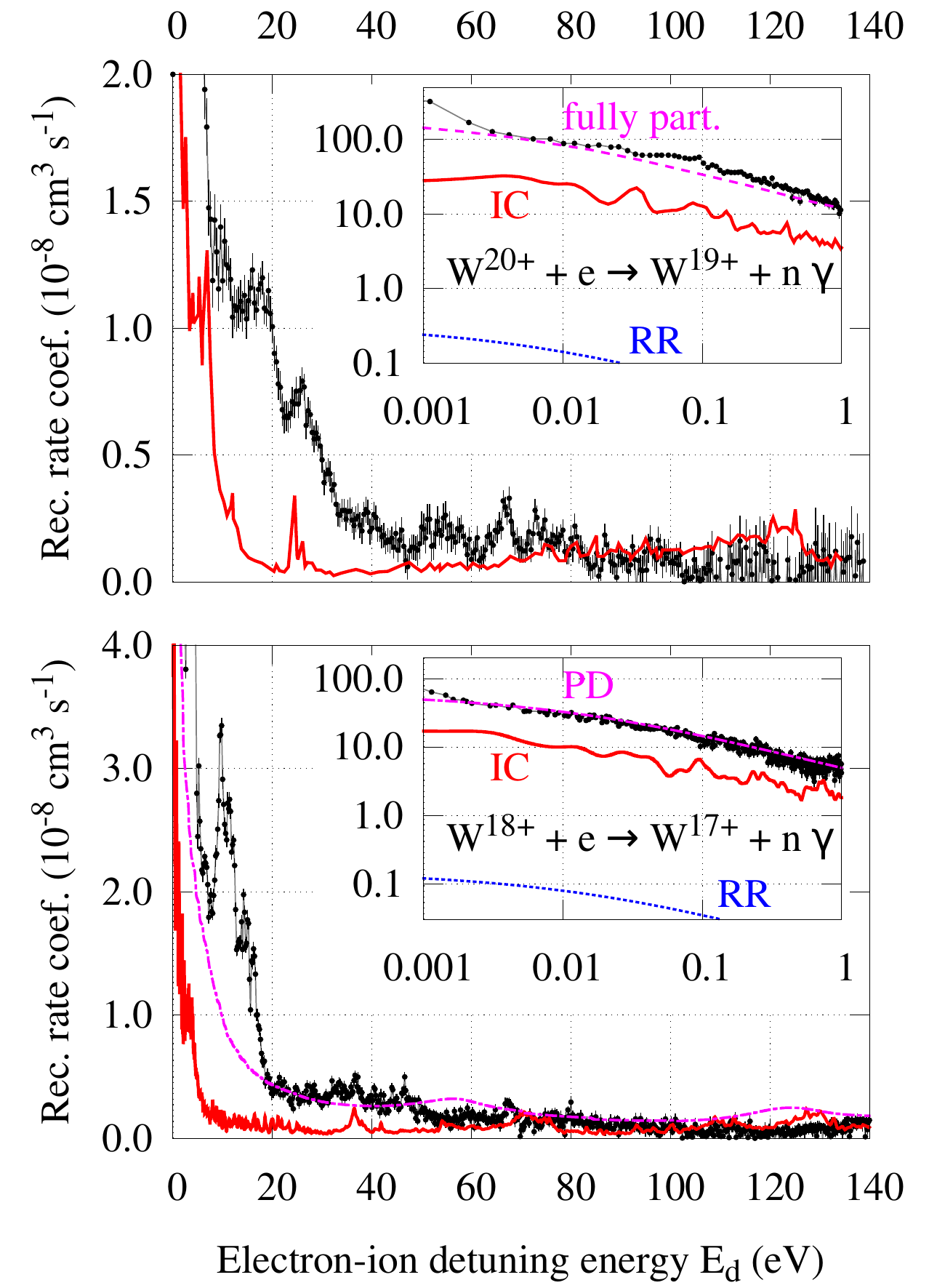}
 \caption{Merged-beams recombination rate coefficients for W$^{20+}$ (top) \cite{schippers2011,badnell2012} and W$^{18+}$ (bottom) \cite{spruck2014}. The energy-resolved experimental data are represented by the black dots in each frame, where the insets show logarithmic blow-ups of the lowest-energy regions. The lines show different theoretical calculations. The blue dotted lines represent the calculated rate coefficient for pure radiative recombination (RR), which falls short of experiment by almost three orders of magnitude. The solid red curves are state-of-the-art intermediate coupling (IC) calculations \cite{badnell2012,spruck2014} based on the \textsc{Autostructure} code, which in case of the open-4$f$ shell tungsten ions, significantly fall below the experimentally observed amplitudes as well. The magenta dashed curve in the W$^{20+}$ plot is the initial `fully-partitioned' calculation, that matches the experiment at low energies, but could not be extended to higher collision velocities \cite{badnell2012}. The latest partitioned-and-damped (PD) model -- shown on the example of W$^{18+}$ as the dash-dotted magenta curve -- matches the experimental signal strength at all energies \cite{spruck2014} (see text).\label{mbrrcfig}}
\end{figure}

Primarily, the experiments yield collision-energy resolved recombination rate coefficients, convolved only with the apparatus resolution of the merged-beams set-up. We refer to these as `merged-beams recombination rate coefficients' (MBRRCs). The data for W$^{20+}$ \cite{schippers2011} and W$^{18+}$ \cite{spruck2014} is depicted in Fig.~\ref{mbrrcfig}.\par
In all experiments, extremely high recombination rate coefficients of the order of $10^{-6}$~cm$^3\,$s$^{-1}$ at near-zero collision energy have been found. These are in fact the highest rates ever observed in recombination of highly-charged atomic ions, and are rivalled only by single-pass experiments on Au$^{25+}$, another ion with open 4$f$ sub-shell, where similarly high values have been reported \cite{hoffknecht1999}.\par 
In all measurements, the low-energy recombination rate coefficients of the tungsten ions are enhanced by almost three orders of magnitude compared to radiative recombination (RR), which can be computed reliably \cite{hoffknecht2000,trzhaskovskaya2013}. Already in the case of Au$^{25+}$, this phenomenon had been interpreted as strong domination of the recombination by resonant processes like dielectronic recombination \cite{hoffknecht1999}. In the tungsten experiments, that interpretation was additionally supported by the fact that all measured MBRRCs present rich structure. The visible features are much broader than the experimental energy resolution as defined by the temperatures of the target electron and stored ion beams,  hence they were attributed to large, unresolved arrays of recombination resonances \cite{schippers2011}.\par
The MBRRCs can be converted into temperature-dependent plasma rate coefficients (PRRCs) by convolution with a Maxwellian electron energy distribution. Results are shown in Fig.~\ref{prrc}. We noted early-on that the recombination rate coefficients obtained from the ADAS database \cite{foster2008,adas} fall far short of the experimental ones \cite{schippers2011}. At very low temperature, the ADAS database rate coefficients approach the value for purely radiative recombination. From this observation it had already been concluded that an important class of resonant processes had so far been neglected in the results \cite{puetterich2008,foster2008} held in the ADAS database. The fact that this leads to discrepancies of up to factors of 20 between experiment and the ADAS database at temperatures relevant for modelling tungsten impurities in tokamak plasma (cf.\ Fig.~\ref{prrc}), has attracted some attention \cite{li2016}.\par
Already in a first heuristic analysis of the measured W$^{20+}$ MBRRC, we speculated that the large number of auto-ionising states below $\sim$50~eV, attached to the available fine-structure levels of the ionic ground configuration, were responsible for the enormously enhanced recombination rates at low collision energies \cite{schippers2011,krantz2014}. It was clear that a proper modelling of the processes would require a treatment within the scope of state-of-the-art atomic structure theory.     
\section{Theoretical description\label{theorychapter}}
It has been known for a long time that pure radiative recombination (RR) calculations are suited only to electron collisions with bare nuclei, and that they can grossly underestimate the recombination rates of ions with a residual electron shell \cite{shore1969}. However, the observed resonant processes had never been as dominant as in the case of the open 4$f$-sub-shell ions. 
\subsection{Resonant recombination}
Dielectronic recombination is initiated by a capture (the inverse Auger process) of an initially free electron by an ion, with simultaneous excitation of a bound electron of the latter \cite{burgess1964,shore1969}. This forms an intermediate, `doubly-excited' auto-ionising level, which subsequently stabilises by photon emission. Auto-ionisation of the transient state competes with the radiative stabilisation channels. Since the capture step is radiationless, DR is a resonant process, leading to enhancements of the recombination cross section at specific electron-ion collision energies.\par
DR of simple atomic systems has been studied extensively, both experimentally and theoretically \cite{wolf2000,beilmann2011}. In electron cooler experiments, DR is typically the dominant recombination process for most ions with simple shell structure, although the higher-order process of tri-electronic recombination has also been clearly identified in Be-like ions \cite{schnell2003}. For very simple Li-like systems, DR calculations based on relativistic many-body perturbation theory (RMBPT) are so accurate that they allow probing of QED effects by comparison to the observed resonance structure in electron target experiments \cite{lestinsky2008,brandau2008}. For ions with larger electronic shells, multi-configuration Dirac-Fock (MCDF) and multi-configuration Breit-Pauli (MCBP) calculations have been found to have good predictive power, as long as the number of active electrons of the systems is such that the total amount of intermediate levels accessible via DR stays numerically manageable \cite{savin2003,schmidt2007}.\par
Our workhorse for atomic structure computations is the \textsc{Autostructure} code, an MCBP implementation \cite{autostructure}. Within that framework, DR is modelled using the independent-processes, isolated-resonances and distorted-waves (IPIRDW) approach \cite{badnell2006}. Therein, the (partial) resonance strength for dielectronic recombination from an initial level $i$ of the parent ion into a defined final level $f$ of the daughter via an auto-ionising level $u$ of energy $E_u$ is given as
\begin{equation}
 \label{ipirdw}
 \hat{\sigma}^u_{f,i}(E_u) = \frac{(2\pi a_0 I_H)^2}{E_u} \frac{\omega_u}{2\omega_i} \frac{ \tau_0\, A^a_{u\rightarrow i}\,A^r_{u\rightarrow f} }{ \sum_{i^\prime} A^a_{u\rightarrow i^\prime}  + \sum_{f^\prime} A^r_{u\rightarrow f^\prime} }\ . 
\end{equation}
$A^a$ and $A^r$ are the rates for auto-ionisation and radiation, respectively. $\tau_0$ is the atomic time unit, $a_0$ the Bohr radius, and $I_H$ the ionisation energy of the hydrogen atom. $\omega_i$ and $\omega_u$ are the statistical weights of the initial and intermediate levels $i$ and $u$, respectively. The sums go over all levels $i^\prime$ and $f^\prime$ of the parent and daughter ion, respectively, that are accessible from $u$ via either auto-ionisation ($A^a$) or radiative decay ($A^r$).\par
The energy-resolved resonance can be obtained from Eq.~\ref{ipirdw} via
\begin{equation}
 \sigma^u_{f,i}(E) = \hat{\sigma}^u_{f,i}(E_u) L^u(E)\ ,
\end{equation}
where $L^u$ is a Lorentzian profile of width $\Gamma_u$ centred on energy $E_u$. However, for practical calculations and comparison to experimental data, one defines an average partial resonance strength in a bin of width $\Delta E$ by   
\begin{equation}
 \label{theores}
 \bar{\sigma}^u_{f,i}(E_u) = \frac{1}{\Delta E} \hat{\sigma}^u_{f,i}(E_u)\ ,
\end{equation}
which is justified as long as $\Delta E$ is much larger than the true resonance width $\Gamma_u$, but smaller than the experimental energy resolution.\par 
Remember that emitted photons are not detected in a typical electron cooler experiment (cf.\ Sect.\ \ref{expsect}) and so neither the final nor the intermediate level leading to a successful recombination is individually observed. Hence, summation over all levels $u$ accessible via capture, and over all levels $f$ below the ionisation threshold must be performed to obtain the total resonance strength in the cooler experiment:
\begin{equation}
 \label{sumdr}
 \bar{\sigma}_i(E_n) = \sum_{u,f} \bar{\sigma}^u_{f,i}(E_u) \quad \forall\ E_u \in [E_n,E_n + \Delta E[ \ .
\end{equation}
Therein, the values $E_n$ correspond to the binning of the energy axis which naturally results from the above chosen resolution $\Delta E$. 
\begin{figure}[tb]
 \centering
 \includegraphics[width=8.3cm]{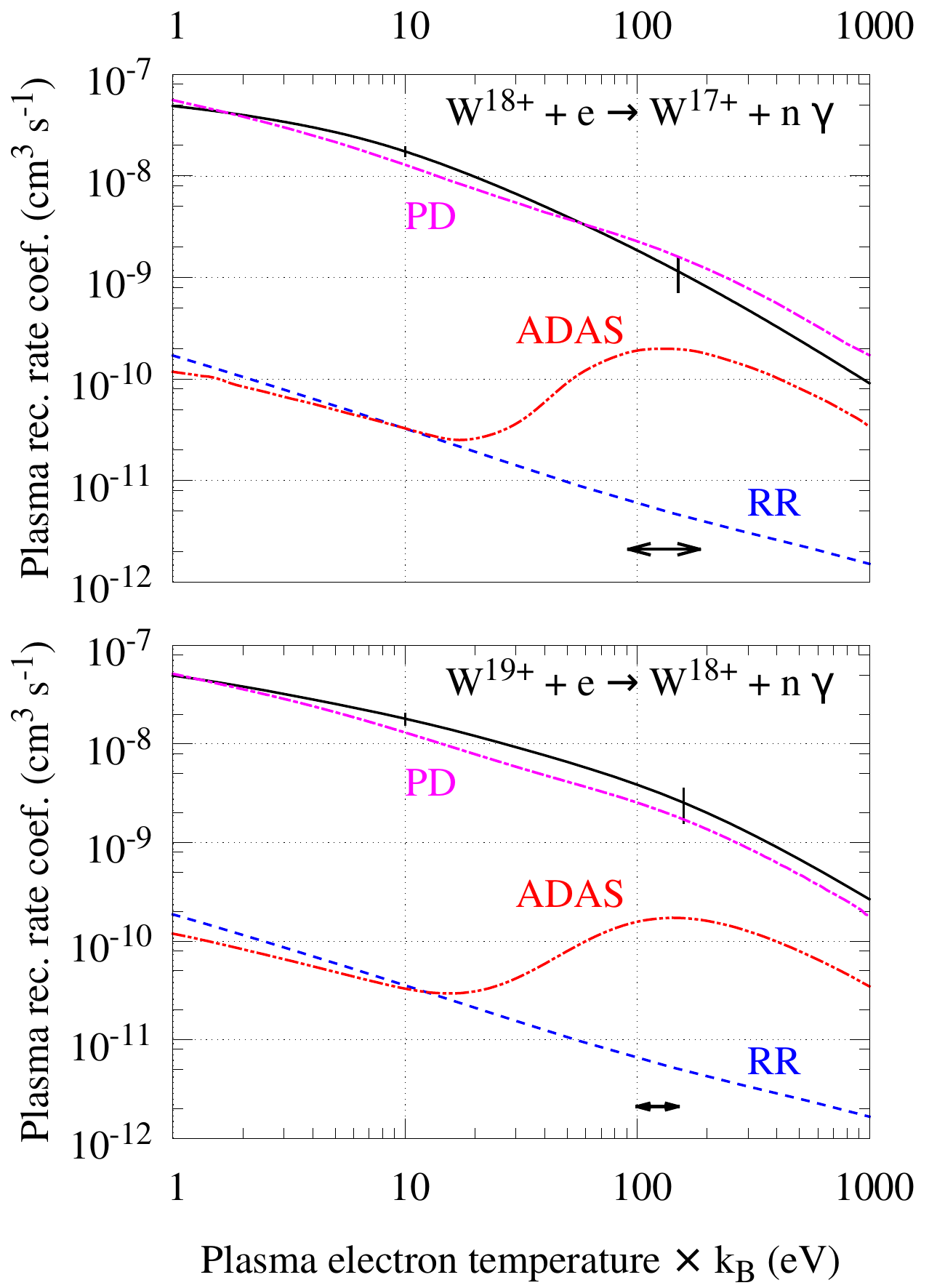}
 \caption{Experimental and theoretical plasma recombination rate coefficients (PRRCs) for W$^{18+}$ (top) and W$^{19+}$ (bottom) \cite{spruck2014,badnell2016}. The experimental data, obtained by convolution of the measured MBRRCs are represented by the solid black curves. The vertical bars indicate the experimental uncertainties. The most recent partitioned-and-damped (PD) calculation is depicted by the magenta dash-dotted lines. The blue dashed lines are the recombination rate coefficient from purely radiative recombination (RR). The red dash-dot-dotted lines represent the PRRC from the ADAS code \cite{adas}. The horizontal arrows indicate the plasma temperature region where the fractional abundance of the respective tungsten charge state peaks.\label{prrc}}
\end{figure}
\subsection{Addressing the open f-shell}
We compute all atomic levels, radiation rates, and auto-ionisation rates using \textsc{Autostructure}. The code, as it existed at the beginning of the tungsten studies, had previously been used successfully to model DR of Au$^{20+}$ with its [Kr]~$4d^{10}4f^{13}$ configuration \cite{ballance2012}. However, adaptation to the `half-open' $4f$ sub-shells of W$^{18+}$ ([Kr]~$4d^{10}4f^{10}$) to W$^{21+}$ ([Kr]~$4d^{10}4f^{7}$) required significant enhancements, as the number of levels that needed to be considered rose drastically \cite{badnell2012}.\par
Of the four tungsten charge states for which experimental recombination rate coefficients had been measured at the TSR, W$^{20+}$ ([Kr]~$4d^{10}4f^{8}$) was the first to be studied theoretically using the new code version. Configuration-averaged (CA), $LS$, and intermediate coupling (IC) schemes were used. The latter is the most computationally demanding but promised to best reproduce the complex-shell situation, as it allows for level mixing \cite{badnell2012}.\par
The result of that calculation is depicted in the top panel of Fig.~\ref{mbrrcfig} (solid red curve), together with the experimental MBRRC for W$^{20+}$ \cite{schippers2011}. For comparison, the theoretical resonance structure is convolved with a typical experimental energy distribution derived from the temperatures of the probing electron beam. One can see that the calculation falls short of the experiment at low collision energies (below $\sim 40$~eV) by a factor of 3--4, while the result labelled `PD' (described next) is in agreement with the experiment.\par 
The missing signal in the calculation is believed to result from incomplete mixing in the IC model compared to nature \cite{badnell2012,spruck2014}. For numerical reasons, the calculation is essentially limited to doubly-excited intermediate configurations, resulting from a single electron promotion in the parent ion with simultaneous capture of the initially free electron. In a very complex shell, featuring a quasi-continuum of states, such a pure doubly-excited configuration will mix with a multitude of neighbouring configurations, including multiply excited ones.\par 
As was first pointed out by Flambaum et al., most of the multiply excited configurations cannot couple to the ground state of the next ionization stage in lowest order perturbation theory, although radiative stabilisation is possible normally \cite{flambaum2002}. Hence, if they mix strongly with 
states that do so couple, they can effectively contribute to an enhanced recombination process.\par
A simple reasoning illustrates this behaviour \cite{badnell2012}: in Eq.~\ref{ipirdw}, consider a level $u$ such that $A^a \ll A^r$. Then it follows in Eq.~\ref{sumdr} that $\sum_f \bar{\sigma}^u_f \sim A^a$. Mixing changes nothing to this situation, as the initial capture rate is the limiting factor in the reaction. However, in the opposite case of a transient level initially with $A^a \gg A^r$, mixing \emph{does} change the overall recombination yield. In absence of mixing we have $\sum_f \bar{\sigma}^u_f \sim A^r$. Now assume that the intermediate level can mix with a set of levels $m$ from multiply excited configurations, which are characterised by very weak capture rates. If the mixing is sufficiently strong, the original capture rate (proportional to $A^a$) is diluted over the states $m$ such that each satisfies $A^a_m \ll A^r$ (assuming that the radiative rate of the mixed set is similar to the unmixed one). Hence, we have $\sum_f \bar{\sigma}^u_f \sim \sum_m A^a_m \approx A^a$ and so the total recombination is enhanced by a factor $A^a/A^r$.\par
We use a Lorentzian (or Breit-Wigner) redistribution \cite{badnell2012,spruck2014} with a width characterising chaotically mixed states \cite{flambaum2002} (typically 10~eV). We find little difference from redistributing or `partitioning' over `physical' multiply excited states, as determined by large-scale configuration-average calculations, say, or simply redistributing uniformly over a set of bin energies \cite{spruck2014}. The reason for this is simply the density of states available in the open $4f$ sub-shell. We note that the former approach appears to be equivalent to the working formula of `statistical' theory \cite{flambaum1994} as applied to DR \cite{flambaum2002}. The idea of partitioning arose originally back in the late 1980s \cite{badnell1989} when attempting to model early measurements on DR at Oak Ridge National Laboratory. Configuration average results were statistically partitioned over levels to try and simulate the parent term splitting observed in measurements on Be-like ions. The weighting function used now is different but the methodology is very similar.\par
The amount of missing recombination signal in the IC calculation for W$^{20+}$ recombination was first estimated by expanding the mixing of the available levels in the \textsc{Autostructure} calculation to an unphysically broad range of energies \cite{badnell2012}. We refer to this calculation as the `fully partitioned' model, which is shown as the dashed curve in Fig.~\ref{mbrrcfig} (top, inset). This leads to a good matching with the experiment at the lowest collision energies. However, this approach is also valid only for that extreme low-energy region: At rising energy of the intermediate levels populated in the collision, the total recombination yield quickly decreases, as more and more auto-ionisation channels into excited states of the parent ion open up.\par
In parallel to our own theoretical study of W$^{20+}$, Dzuba et al.\ published initial calculations on the same and the neighbouring ion charge states, in which they used a statistical framework for the configuration mixing \cite{dzuba2012}. On that basis, they soon afterwards presented an updated theory for W$^{20+}$, which also correctly implemented the reduction of the total recombination yield at higher collision energies. Using that model, they were able to reproduce the experimentally observed amplitude of the MBRRC at all energies \cite{dzuba2013}.\par  
In the subsequent theoretical studies on W$^{18+}$ ([Kr]~$4d^{10}4f^{10}$) and W$^{19+}$  ([Kr]~$4d^{10}4f^{9}$), we also extended our own calculations to include the reduction of the recombination rate coefficient at high energies. This is referred to as the partitioned-and-damped (PD) model. It is shown (magenta dash-dotted curves) in the lower half of Fig.~\ref{mbrrcfig} for the case of W$^{18+}$, together with the measured data from the corresponding TSR experiment. One sees that the PD model describes the experimental signal strength well over the entire energy range under study. Some prominent resonant features of the measurement are not reproduced. This is hardly surprising, as obviously-important (multiply-excited) configurations are omitted from the detailed IC calculations as discussed. Consistently with the previous findings on W$^{20+}$, the non-partitioned IC calculation (solid red curves) falls short of the experimental data, by a factor $\sim 3$.\par 
Our recently published study on W$^{19+}$ showed similarly good agreement in the experimental and theoretical MBRRCs \cite{badnell2016}.
\subsection{Effects of metastable ion populations}
\begin{figure}[tb]
 \centering
 \includegraphics[width=8cm]{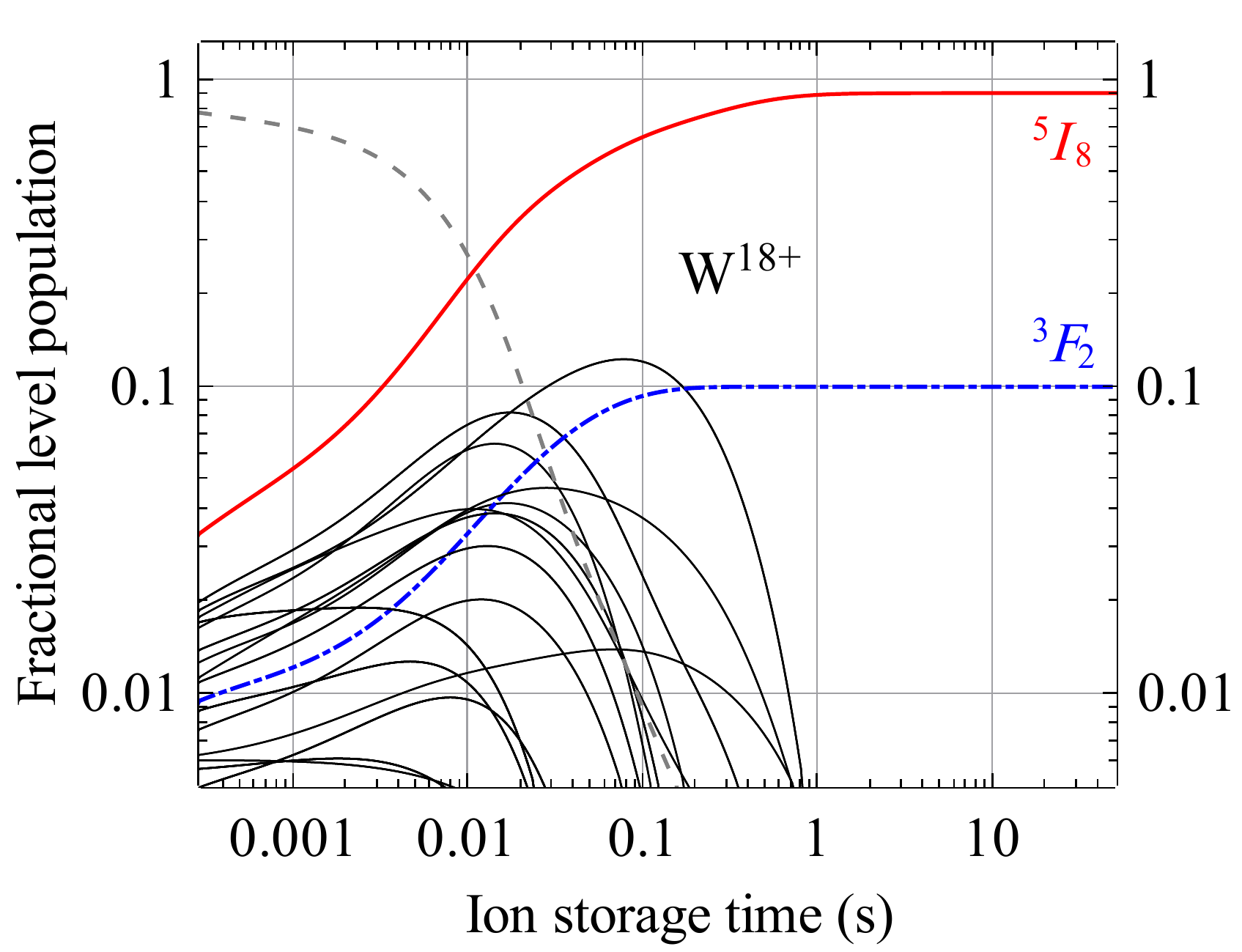}
 \caption{Example of imperfect beam-cleaning by spontaneous decay. The plot shows the calculated evolution of the W$^{18+}$ metastable level population after ion production by electron stripping \cite{spruck2014}. The fine black lines show the fractional abundance of the 17 longest lived metastable levels. After 1~s of storage time, i.e.\ at the beginning of the experiment, 90\,\% of the ions have decayed to the $^5I_8$ ground level (solid red line). However, the remaining 10\,\% of ions accumulate in the metastable level $^3F_2$ (blue dash-dotted line), that is so long-lived that it cannot decay during the storage time in the TSR. The grey dotted line shows the sum fractional population of all other (1652) levels of the ground and first excited configurations of W$^{18+}$. \label{w18levels}}
\end{figure}
It has been first pointed out for W$^{20+}$ \cite{schippers2011} that the high level of complexity of the open $4f$ sub-shell of the ions under study, combined with the limited storage time of the ions in the TSR, may lead to complications regarding the initial-level population that needs to be considered in the analysis.\par
In the ion-production process, involving high-energy electron stripping in the linear accelerator, a large amount of low-lying (a few tens of eV) fine-structure levels of the ground configuration are accessible for all tungsten charge states under study \cite{spruckphd,krantz2014}. The majority of these have natural lifetimes that are short enough to allow them to decay during the beam preparation phase (1--2~s) in the storage ring. However, for each ion, a few levels are predicted to be so long-lived that their population in the experiment cannot be neglected. E.g., for W$^{18+}$ the excited level $^3F_2$ was predicted to have a fractional abundance of 10\,\% during the data taking, whereas the remaining 90\,\% of ions would populate the $^5I_8$ ground level (cf.\ Fig.~\ref{w18levels}) \cite{spruck2014}. A non-negligible metastable ion population can make comparison between theory and experiment difficult, since computations of dielectronic recombination are done traditionally assuming a ground-level population of the parent ion.\par
The problem was finally studied thoroughly for the case of W$^{19+}$. There, a total of 36\,\% of the ions were predicted to populate four different metastable levels, whereas only two-third of the ions could be assumed to have been in ground-level states. Using our partitioned-and-damped (PD) intermediate-coupling model, we calculated energy-resolved recombination rate coefficients for the 16 lowest excited levels \cite{badnell2016}. Over the studied collision energy range the results varied only within 10\,\% in amplitude, which is within the experimental uncertainty of the measurements.\par 
The fact that there appears to be little dependence of the recombination rate coefficient on the parent excitation is not too surprising. While for simple systems, DR from excited levels is known to be significantly suppressed by auto-ionisation into lower-lying levels of the ion, in the scenario considered here, following complete redistributive mixing, capture from the ground and low-lying metastable levels have access to similar suppression. This result is encouraging when it comes to comparisons of the storage ring measurements to theory as well as to observations of thermal plasma.             
\subsection{Comparison of theory and experiment for plasmas}
Similar to the experimental MBRRC data, the theoretical recombination cross sections can be convolved with a thermal electron energy distribution to yield plasma recombination rate coefficients (PRRCs). The results for W$^{18+}$ and W$^{19+}$, which mark the present state-of-the art of our theoretical research, are shown in Fig.~\ref{prrc} \cite{spruck2014,badnell2016}.\par
The PD theory described above matches the experiment within its uncertainty over the entire plasma temperature range under study. Compared to the ADAS database, the improvement in the calculated PRRCs is spectacular. This is especially true at the lowest thermal energies $k_BT \le 10$~eV, where the ADAS database rate coefficients \cite{foster2008} are unable to reproduce the resonant enhancement of the recombination rate coefficient with respect to pure RR. However, also in the temperature ranges around 100--200~eV, where the charge states under study should have their highest abundance in an electron collision dominated plasma, the advantage of the PD theoretical description is significant.\par
Finally, the good level of agreement with our experimental MBRRC achieved by Dzuba et al.\ \cite{dzuba2013} suggests that convolution of their data to produce PRRCs would give an accuracy similar to our own calculations. 
\section{Summary and conclusion\label{discussion}}
As part of a wider research campaign targeting all electron collision reactions of highly-charged tungsten ions \cite{mueller2015}, the studies we have presented in this review had two goals. The first was to establish a sound experimental benchmark for the electron recombination of tungsten ions in the range of charge states characterised by an open 4$f$ sub-shell. Through our electron-target measurements on stored and cooled W$^{18+}$, W$^{19+}$, W$^{20+}$, and W$^{21+}$ at the Heidelberg TSR storage ring \cite{schippers2011,spruck2014,badnell2016,spruckphd} this part of the endeavour has been achieved. The very object under study -- namely the enormously enhanced recombination rate coefficient of these ions -- rendered the experiments quite difficult, and the fact that they could be performed successfully is an indicator for the high versatility of the TSR and for the great skill of its accelerator crew.\par
The second goal was to develop a theoretical model that can describe the recombination of the ions under study. Here, good progress has been made, both, by introduction of our own partitioned-and-damped approach \cite{badnell2012,badnell2016}, and by the equivalent `statistical' approach of Dzuba and coworkers \cite{dzuba2012,dzuba2013}. A pure, {\it a-priori}, intermediate coupling calculation within the MCBP framework turned out to be incapable of fully reproducing the measured resonant enhancement of the recombination rate coefficients at low energies. The reason lies in the fact that it is at present technically impossible to model the full complexity of the `half-open' 4$f$ sub-shell configuration numerically. In this situation, it is unsurprising that the present IC calculations also fail to reproduce important resonant features that are observed in the experimental data. We must therefore admit that theory is not yet able to provide the same degree of level-to-level understanding of the recombination process that is achieved in DR of more simple atomic ions. It is not clear yet, how this limitation of the {\it ab-initio} calculations can be overcome.\par
In this situation, the above described partitioned-and-damped model has proven to provide a way to predict the total amount of missing recombination yield. On the basis of the excellent agreement of our theoretical PRRCs with the experiment, we strongly encourage the use of these results over the simpler ones held in the ADAS database.\par
Beyond their immediate relevance to fusion plasma modelling, the here featured studies are among the few detailed investigations on quantum dynamics of atomic systems of extreme complexity. Already in their work on recombination of Au$^{25+}$ -- which evolved into the successful `statistical' model of W$^{20+}$ recombination provided by Dzuba et al.\ \cite{dzuba2012,dzuba2013} -- Flambaum et al.\ noted that their `chaotic-mixing' ansatz may be a recipe for numerical treatment of a much wider range of phenomena. Besides atoms with large electronic configurations, they suggested complex systems of nuclear, molecular, and solid-state physics -- all characterised by dense spectra of multiply-excited states -- as possible applications \cite{flambaum2002}. In a recent follow-up publication, they elaborate on these ideas \cite{flambaum2015}. 
\section*{Acknowledgements}
We thank all our co-workers in this project, especially the accelerator operations team of the Max Planck Institute for Nuclear Physics for their support in the experiments. We gratefully acknowledge support by the German Research Foundation (DFG) under contract numbers Mu-1068/20 and Schi-378/9, as well as support to the University of Strathclyde by UK EPSRC grant EP/L021803/1.

\end{document}